\documentclass[journal]{IEEEtran}

\usepackage[utf8]{inputenc}
\usepackage{graphics,graphicx,color,epsfig}
\usepackage{amsmath,amssymb,amsthm,amsfonts,bm,bbm}
\usepackage{algorithmicx,algorithm,algpseudocode}
\usepackage{multirow}
\usepackage{cite}
\usepackage[normalem]{ulem}

\usepackage{tikz}
\usetikzlibrary{automata, positioning, arrows,shapes.multipart,fit}

\usepackage{todonotes}
\presetkeys{todonotes}{color=blue!20}{}

\newcommand{\revDel}[1]{}

\usepackage[normalem]{ulem}

\title{Realizing Neural Decoder at the Edge with Ensembled BNN}
\author{Devannagari Vikas, Nancy Nayak, and Sheetal Kalyani\\
Department of Electrical Engineering, Indian Institute of Technology Madras, India. \\
Emails: \{ee19m018@smail,ee17d408@smail,skalyani@ee\}.iitm.ac.in
}
\begin{document}
\maketitle
	\begin{abstract}
		In this work, we propose extreme compression techniques like binarization, ternarization for Neural Decoders such as TurboAE. These methods reduce memory and computation by a factor of $64$ with a performance better than the quantized (with $1$-bit or $2$-bits) Neural Decoders. However, because of the limited representation capability of the Binary and Ternary networks, the performance is not as good as the real-valued decoder. To fill this gap, we further propose to ensemble $4$ such weak performers to deploy in the edge to achieve a performance similar to the real-valued network. These ensemble decoders give $16$ and $64$ times saving in memory and computation respectively and help to achieve performance similar to real-valued TurboAE.
	\end{abstract}
	
	\section{Introduction}
The future wireless communication system 6G will not only be equipped with multi-band high-speed transmission but also energy-efficient communication, low latency, and high security. In a digital communication system, different physical layer encryption algorithms like LDPC, Polar, Turbo codes \cite{LDPCmackay1996near, Turboberrou1993near} are used as channel coding methods\cite{dehdashtian2021deep} to prevent the data from getting corrupted by channel noise. 
When the channel deviates from the Gaussian setting in a practical scenario, to exploit the power of the encoder, Neural Networks (NN) have been used to design the decoder while the encoder is fixed as a near-optimal code \cite{doan2018neural}. Deploying decoders for these codes takes up huge computation which is only possible because of recent advancements in signal processing methods. With a surge in the number of devices in the network, the interactions among themselves may result in excessive signal processing at the user end that gives rise to huge power consumption. Therefore economic energy usage to have elongated battery life in mobile devices has been a research direction of utmost importance\cite{zhan2018state,mahapatra2015energy,nayak2020green}. In a noisy channel, encoding has been challenging even though the decoders have good performance\cite{vedula2020joint}; so the authors in \cite{raj2018backpropagating,o2017deep, zhu2019joint, raj2020design} proposed neural code where the encoder and decoder are jointly trained. To overcome the problem of convergence to a local minimum in joint optimization, \cite{NEURIPS2019_228499b5} proposed TurboAE that uses Convolutional Neural Network (CNN) based over-complete Auto Encoder (AE) model incorporating interleavers and de-interleavers to achieve the performance of State Of The Art (SOTA) channel codes under the AWGN scenario. All the existing neural AEs have real-valued network parameters and perform floating-point operations during deployment. For instance, the TurboAE architecture has nearly $26e5$ parameters that take up a memory of $20.84$ MB considering a $64$ bit floating-point representation. Out of total $26e5$ parameters, the encoder has nearly $1.5e5$ parameters whereas the decoder has nearly $25e5$ parameters. Because of the huge number of parameters in the AEs, deploying it in a resource-limited Internet Of Thing (IoT) setup is a challenging task. Furthermore, with the advent of edge computing in IoT scenarios, the computation is decentralized to edge devices where the data is processed locally. Realizing a Neural Decoder such as TurboAE\cite{NEURIPS2019_228499b5} at a user end that has limited memory and computing power is not practically feasible. 




\subsection{Contributions}
In the domain of wireless communication, the channel noise is real-valued and till now, only the real-valued Neural Decoders have been used for end-to-end training and these use only floating-point operations. In this work, we have explored the possibilities of using extreme compactification techniques in Machine Learning-based wireless decoders like TurboAE. We further propose techniques that allow the decoder to be memory and computation-efficient but still have a performance close to the real-valued decoder. The major contributions of our work are the following: 
\begin{enumerate}
    \item We propose to use binary filters/weights/biases and binary activations\footnote{Binary Neural Networks\cite{NIPS2016_d8330f85} take the compression to the extreme level by replacing $64$ bit floating point (FP) weights and activations to be $1$-bit that gives a memory reduction of $64$ times. Also the FP multiplication and addition operations are replaced with \textit{xnor} and \textit{popcount} operations that reduces computation cost radically during the inference time.} in the Neural Decoder to save in memory and computation at the edge. 
    \item The performance is further improved by the use of a Ternary Neural Network (TNN) where the weights take three levels $\{-1,0,+1\}$ with the binary activation. The proposed architectures with binary and ternary weights are shown to be better than one where the trained network is quantized with $2$ bit or $1$ bit. 
    \item An ensemble of multiple weak binary and ternary decoders is then proposed and is shown to perform close to the real-valued TurboAE and also achieve a $16$ times saving in memory and nearly $64$ times speed up due to less computation thus enabling us to achieve energy efficiency and low latency in the edge communication. 
\end{enumerate}

Before discussing different compressed versions of TurboAE, we first review the extreme compression techniques like BNN and TNN in Sec. \ref{sec:BNN} and then study the impact of these techniques in TurboAE in Sec. \ref{sec:TurboAE}. 

	\section{Extreme compression techniques}\label{sec:BNN}
We denote a real valued NN $g_{\boldsymbol{\phi}}(.)$ where $\boldsymbol{\phi}$ represents the real valued network parameters. The output from the NN is given by $\mathbf{y}=g_{\boldsymbol{\phi}}(\mathbf{x})$ where $\mathbf{x}$ is the input features to the NN and can be real valued. The neural network $g_{\boldsymbol{\phi}}(.)$ can be of any type: a fully connected, a CNN or a Recurrent Neural Networks (RNN). As TurboAE uses a CNN for the Neural Decoder, we now focus on CNNs. For $g_{\mathbf{\phi}}(.)$ a CNN of $L$ layers, the parameters are the filters of the CNN and are given by $\mathbf{\phi}=\{\mathbf{W}_1,\dots,\mathbf{W}_L\}$ where $\mathbf{W}_l\in \mathbb{R}^{c_o \times c_i \times k}$ for $l^{th}$ layer of one dimensional CNN. Here $c_i$ and $c_o$ represents number of input and output channels and $k$ is the dimension of the filter. For a one dimensional CNN as used in TurboAE, if the input to $l^{th}$ layer of CNN has spatial features of dimension $h_{in}$, then input to $l^{th}$ layer is $\mathbf{a}_l \in \mathbb{R}^{c_i\times h_{in}}$. The output of $l^{th}$ layer is $\mathbf{a}_{l+1}\in\mathbb{R}^{c_{o}\times h_{out}}$ where $h_{out}$ is the dimension of the output. For a Binary Neural Network (BNN), the weights and activations ($\mathbf{W}$ and $\mathbf{a}$) are binarized using the $sign$ function before taking convolution.
\begin{align}
    b = sign(r)=\begin{cases}
    +1, & \text{if } r \geq 0 \\
    -1, & \text{otherwise}.
    \end{cases}
\end{align} 
The binarized parameter $\mathbf{W}^b_l$ and $\mathbf{a}^b_l$ is given by:
\begin{align}
    \mathbf{W}^b_l = sign(\mathbf{W}_l), \text{ and }
    \mathbf{a}^b_l = sign(\mathbf{a}_l)
\end{align}
The real-valued convolution is approximated with binary weights and activations as $\mathbf{W}_l\ast\mathbf{a}_l\approx \mathbf{W}^b_l\circledast \mathbf{a}^b_l$ where $\circledast$ is convolution performed with bitwise operations. Even though the binarized weights are used for the forward pass, only the real-valued latent weights are updated with the real-valued gradients during backpropagation. However, during inference, these latent weights can be dropped and a binary network with the binary weights and activations can be deployed. The $sign$ function is non-differentiable and has gradients as zero almost everywhere; thus it is not appropriate for the backpropagation during the training. Therefore a straight-through estimator \cite{bengio2013estimating} was proposed that binarizes in the forward pass but during backpropagation, it just passes the gradients as it is to the previous layers. For instance, if $b=sign(r)$, then $grad_r=grad_b\mathbf{1}_{|r|\leq1}$ where $grad_r=\frac{\partial C}{\partial r}$, $grad_b=\frac{\partial C}{\partial b}$ and $C$ is the cost function of the NN. To have a stable update during the training, the updated real-valued weights are clipped between $[-1,1]$. 

If a real-valued network $g_{\boldsymbol{\phi}}(.)$ is deployed in a $64$ bit system, then its binary version will occupy $64$ times lesser memory and all the floating-point operations can be converted to just xnor and popcount operations. However, because of this extreme compactification, the performance generally degrades significantly. So \cite{li2016ternary} proposed to use Ternary Neural Network (TNN) where $3$ bits $\{-1,0,1\}$ are used. Therefore the ternarized parameter $t$ is given by:
\begin{align}
    t = tern(r) = \begin{cases}
+1, & \text{if } r > \Delta \\
0, & \text{if } r < |\Delta| \\
-1, & \text{if } r< -\Delta.
\end{cases}
\end{align}
where $\Delta \approx 0.7 E(|\mathbf{r}|)$ in our architecture where $\mathbf{r}$ the set parameters of the real network. The introduction of zero as another bit along with $\{+1,-1\}$ gives a better representation power and therefore better performance than BNN. But the zero weights need not to be saved during deployment. So the memory requirement of TNN is same as that of the BNN. Note that the activation is still binary and thus the computational complexity is also same as the BNN. Therefore with TNN, an improvement in performance over BNN is possible without any degradation in memory requirement or computation.

\begin{figure*}[!t]
    \centering
    \caption{TurboAE interleaved encoder (left), Channel (middle) and TurboAE iterative decoder (right) with block rate $\frac{1}{3}$. $g_{\phi}^v=g_{\phi}^b$ for BinTurboAE and $g_{\phi}^v=g_{\phi}^t$ for TernTurboAE. Fig courtesy \cite{NEURIPS2019_228499b5}}
    \newcommand{\arrowIn}{
\tikz \draw[-stealth] (-1pt,0) -- (1pt,0);
}

\begin{tikzpicture}
 
 \draw[red,dashed]  (0.5,1.25) rectangle (2.75,4) node[above] at (1.625,4) {Encoder $f_{\theta}$};
 \draw[red,dashed]  (4.5,1.25) rectangle (16.75,4) node[above] at (10.125,4) {Decoder $g_{\phi}$};

 \draw (0,2.5) -- (0.25,2.5) node[pos=0.5,above]{$\mathbf{u}$};
 \draw (0.25,1.75) -- (0.25,3.25);
 
  \draw (1.5,3) rectangle (2.5,3.5);
 \draw (0.25,3.25) -- (1.5,3.25) node[pos=1]{\arrowIn} node[pos=1.4]{$f_{1,\theta}$};
 \draw (2.5,3.25) -- (3,3.25) node[pos=0.4,above]{$\mathbf{x}_1$} node[pos=1]{\arrowIn}  ;

 \draw (1.5,2.25) rectangle (2.5,2.75);
 \draw (0.25,2.5) -- (1.5,2.5) node[pos=1]{\arrowIn} node[pos=1.4]{$f_{2,\theta}$} ;
 \draw (2.5,2.5) -- (3,2.5) node[pos=0.4,above]{$\mathbf{x}_2$} node[pos=1]{\arrowIn}  node[rotate = 90, pos=1.9]{\small AWGN Channel} ;

 \draw (0.25,1.75) -- (0.75,1.75) node[pos=1]{\arrowIn} node[pos=1.5]{$\pi$};
 \draw (0.75,1.5) rectangle (1.25,2);
 \draw (1.25,1.75) -- (1.5,1.75) node[pos=1]{\arrowIn} node[pos=2.75]{$f_{3,\theta}$};
 \draw (1.5,1.5) rectangle (2.5,2);
 \draw (2.5,1.75) -- (3,1.75) node[pos=0.4,above]{$\mathbf{x}_3$} node[pos=1]{\arrowIn} ;
 
 \draw (3,1.25) rectangle (3.9,3.75);
 \draw (3.9,3) -- (4.5,3) node[pos=0.4,above]{$\mathbf{z}_1$} node[pos=1]{\arrowIn};
 \draw (3.9,2.5) -- (4.5,2.5) node[pos=0.4,above]{$\mathbf{z}_2$} node[pos=1]{\arrowIn};
 \draw (3.9,2) -- (4.5,2) node[pos=0.4,above]{$\mathbf{z}_3$} node[pos=1]{\arrowIn};

 \draw[blue,dashed] (5.25,1.5) rectangle (9.95,3.5) node[above] at (7.7,3.5) {$1^{st}$ iteration};

 \draw (5,3) -- (5.5,3) node[pos=0.3,above]{$\mathbf{z}_1$} node[pos=1]{\arrowIn};
 \draw (5,2.5) -- (5.5,2.5) node[pos=0.3,above]{$\mathbf{z}_2$} node[pos=1]{\arrowIn} node[pos=2]{$g_{\phi_{1,1}}^{v}$};
 \draw (5,2) -- (5.5,2) node[pos=0.3,above]{$p_0$} node[pos=1]{\arrowIn};
 \draw (5.5,1.75) rectangle (6.4,3.25);
 
 \draw (6.4,2) -- (6.75,2) node[pos=0.4,above]{q} node[pos=1]{\arrowIn} node[pos=1.8]{$\pi$};
 \draw (6.75,1.75) rectangle (7.2,2.25);

 \draw (7,3) -- (7.7,3) node[pos=0.3,above]{$\pi(\mathbf{z}_1)$} node[pos=1]{\arrowIn};
 \draw (7,2.5) -- (7.7,2.5) node[pos=0.3,above]{$\mathbf{z}_3$} node[pos=1]{\arrowIn} node[pos=1.7]{$g_{\phi_{1,2}}^{v}$};
 \draw (7.25,2) -- (7.7,2) node[pos=0.3,above]{p} node[pos=1]{\arrowIn};
 \draw (7.7,1.75) rectangle (8.6,3.25);
 
 \draw (8.6,2) -- (8.95,2) node[pos=0.4,above]{q} node[pos=1]{\arrowIn} node[pos=2.3]{$\pi^{-1}$};
 \draw (8.95,1.75) rectangle (9.7,2.25);
 \draw (9.7,2) -- (9.95,2) node[pos=0.5]{\arrowIn};
 
 \draw [dashed](9.95,2) -- (10.45,2);
 \draw [dashed](9.95,2.5) -- (10.45,2.5);
 \draw [dashed](9.95,3) -- (10.45,3);

 \draw[blue,dashed] (10.5,1.5) rectangle (15.05,3.5) node[above] at (12.875,3.5) {$M^{th}$ Iteration};
 
 \draw (10.25,3) -- (10.75,3) node[pos=0.3,above]{$\mathbf{z}_1$} node[pos=1]{\arrowIn};
 \draw (10.25,2.5) -- (10.75,2.5) node[pos=0.3,above]{$\mathbf{z}_2$} node[pos=1]{\arrowIn} node[pos=2]{$g_{\phi_{6,1}}^{v}$};
 \draw (10.25,2) -- (10.75,2) node[pos=0.3,above]{p} node[pos=1]{\arrowIn};
 \draw (10.75,1.75) rectangle (11.65,3.25);
 
 \draw (11.65,2) -- (12,2) node[pos=0.4,above]{q} node[pos=1]{\arrowIn} node[pos=1.8]{$\pi$};
 \draw (12,1.75) rectangle (12.5,2.25);

 \draw (12.25,3) -- (12.85,3) node[pos=0.3,above]{$\pi(\mathbf{z}_1)$} node[pos=1]{\arrowIn};
 \draw (12.25,2.5) -- (12.85,2.5) node[pos=0.3,above]{$\mathbf{z}_3$} node[pos=1]{\arrowIn} node[pos=1.8]{$g_{\phi_{6,2}}^{v}$};
 \draw (12.5,2) -- (12.85,2) node[pos=0.4,above]{p} node[pos=1]{\arrowIn};
 \draw (12.85,1.75) rectangle (13.75,3.25);
 
 \draw (13.75,2) -- (14.1,2) node[pos=0.4,above]{q} node[pos=1]{\arrowIn} node[pos=2.2]{$\pi^{-1}$};
 \draw (14.1,1.75) rectangle (14.85,2.25);
 \draw (14.85,2) -- (15.2,2) node[pos=1]{\arrowIn} node[pos=2.75]{Sigmoid}; 
 \draw (15.2,1.75) rectangle (16.5,2.25);
 
\draw (16.5,2) -- (17,2) node[pos=0.75,above]{$\hat{\mathbf{u}}$} node[pos=1]{\arrowIn};

\end{tikzpicture}  
    \label{fig:TurboAE}
\end{figure*}
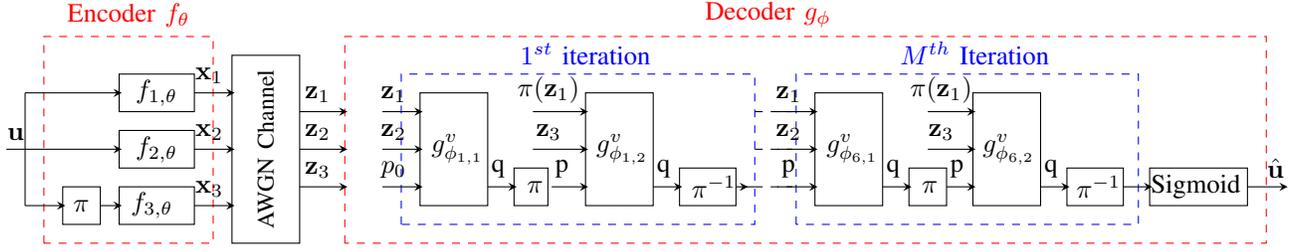

\subsection{Saving in computation}
The convolution between real-valued $\mathbf{W}_l\in \mathbb{R}^{c_o \times c_i \times k}$ and $\mathbf{a}_l \in \mathbb{R}^{c_i\times h_{in}}$ at $l^{th}$ layer results in an output $\mathbf{a}_{l+1}\in\mathbb{R}^{c_o\times h_{out}}$. The total number of multiplication for $l^{th}$ layer is $c_i\times k \times h_{out} \times c_o$ and the total number of addition for $l^{th}$ layer is $(c_i-1)\times (k-1) \times h_{out} \times c_o$. The total count of FLoating Point Operations (FLOP) for $l^{th}$ layer of a real-valued 1D-CNN is the summation of the number of multiplication and addition that is roughly twice of the number of multiplication given by $2\times c_i\times k \times h_{out} \times c_o$.  For a binary counterpart, as the weights and activations are constrained to $-1$ or $+1$, the $64$ bit floating point multiply-accumulation operations are replaced by $1$ bit xnor-count operations \cite{NIPS2016_d8330f85}. Note that the modern CPUs can perform a single multiplication and addition in a single clock cycle, and thus the total number of operations in a binary network is $c_i\times k \times h_{out} \times c_o$. In recent CPUs, $64$ such binary operations can be performed in a single clock cycle hence, giving a speedup of nearly $64$ times in a binary or ternary network \cite{rastegari2016xnor}. Because the filters take only $+1$ or $-1$, only a limited number of filters are possible. So with BNN, the filter repetition can be exploited by using dedicated hardware/software. The implementation on GPU can be made faster by using SIMD within a register (SWAR) where $64$ binary variables are concatenated in a $64$ bit register and a $64$ times speedup on the bitwise operation like xnor can be achieved.

\section{TurboAE and its binarized versions} \label{sec:TurboAE}
The method of channel coding in TurboAE can be divided into three sub-problems: an encoder $f_{\theta}(.)$ at the transmitter, a channel $c(.)$ and a decoder $g_{\phi}(.)$ at the receiver. In a communication system, the encoder $x=f_{\theta}(u)$ encodes the binary bits $\mathbf{u}=(u_1,\dots,u_K)\in \{+1,-1\}^K$ of block length $K$ to get the codeword $\mathbf{x}=(x_1,\dots,x_N)$ of length $N$ such that the codeword satisfies the power constraints. The code rate is $R = \frac{K}{N}$, where $N > K$. The i.i.d. AWGN channel corrupts the encoded bits and generates $z_i = x_i+w_i$ such that $w_i\sim\mathcal{N}(0,\sigma^2)$ for $i=1,\dots,K$. The noise in the AWGN channel is represented by the signal to noise ratio $\text{SNR}=-10\log_{10}\sigma^2$. After transmission through the channel, the decoder $g_{\phi}(z)$ receives the real valued noisy encoded bits $z$ and map them to an estimate of the actual message sequence $\hat{\mathbf{u}}=(\hat{u}_1,\dots,\hat{u}_K)\in \{+1,-1\}^K$ using a decoding algorithm. Channel coding aims to minimize the Bit Error Rate (BER) or the BLock Error Rate (BLER) of the recovered message signal $\hat{u}$ given by $BER=\frac{1}{K}\sum_{1}^{K}Pr(\hat{u}_i\neq u_i)$ and $BLER=Pr(\hat{\mathbf{u}}\neq \mathbf{u})$. Naively applying deep learning models by replacing encoder and decoder with general purpose neural networks does not perform well. So in \cite{NEURIPS2019_228499b5}, authors have proposed a TurboAE with interleaved encoding and iterative decoding using 1D convolutional neural networks. To make the Neural Decoder utilizable at the edge, we first propose to \textbf{\textit{binarize}} and \textbf{\textit{ternarize}} the iterative decoder of TurboAE and inspect its performance. We briefly describe the TurboAE architecture before discussing the proposed compressing techniques.

Turbo code is one of the first capacity approaching codes based on recursive systematic convolutional (RSC) code that has an optimal decoding algorithm namely the Bahl-Cocke-Jelinek-Raviv (BCJR)\cite{bahl1974optimal}. To add long-range memory to the code, interleaving is used: out of two copies of input bits, the first one passes through the RSC code and the second goes through the interleaver before passing through the same RSC code as shown in Fig. \ref{fig:TurboAE}(left). After the transmission through the channel, this code is then decoded by repeating (i) and (ii) alternatively: (i) soft decoding based on the signal received from the first copy (ii) using the de-interleaved version as a prior for decoding the second copy as shown in Fig. \ref{fig:TurboAE}(right). This iterative decoding method keeps re-estimating the posterior distribution on the transmitted bits. Both the interleaved encoder and the iterative decoder are learnable as proposed in TurboAE \cite{NEURIPS2019_228499b5}. The interleaver $\mathbf{x}^{\pi}=\pi(\mathbf{x})$ and the de-interleaver $\mathbf{x}=\pi^{-1}(\mathbf{x}^{\pi})$ shuffles and un-shuffles the input sequence with a random interleaving array known to both encoder and decoder respectively. A code rate of $1/3$ is considered for the interleaved encoder $f_{\theta}$ that has three learnable blocks $f_{1,\theta},f_{2,\theta}$ and $f_{3,\theta}$. The first two takes the original message bit $\mathbf{u}$ to produce $\mathbf{x}_1$ and $\mathbf{x}_2$ whereas the third block takes the interleaved message $\pi(\mathbf{u})$ to return $\mathbf{x}_3$ as shown in Fig. \ref{fig:TurboAE}. The encoded messages are transmitted through the channel and the received noisy messages are $\mathbf{z}_1$,$\mathbf{z}_2$ and $\mathbf{z}_3$. Our focus is mostly on the compression of the iterative decoder part so that it can be deployed at the edge devices. Thus we do not discuss much on the encoder part in this work. Interested readers may refer to \cite{NEURIPS2019_228499b5} for more details on the encoder.

\subsection{Binary and Ternary iterative decoder}
Considering $M(=6)$ iterations of the iterative decoder, each iteration consists of two decoders. First decoder $g_{\phi_{i,1}}(.)$ in $i^{th}$ iteration takes the original noisy message $\mathbf{z}_1,\mathbf{z}_2$ and the prior distribution $p$ on the transmitted bits and returns a posterior $q$ that goes to the second decoder $g_{\phi_{i,2}}(.)$ via interleaving along with the interleaved noisy messages $\pi(\mathbf{z}_1)$ and $\mathbf{z}_3$. In the proposed binarized and ternarized TurboAE, named as \textit{BinTurboAE} and \textit{TernTurboAE} respectively, the real-valued decoders $\{g_{\phi_{1}}, \dots, g_{\phi_{M}}\}$ are replaced with binary decoders $\{g^b_{\phi_{1}}, \dots, g^b_{\phi_{M}}\}$ and ternary decoders $\{g^t_{\phi_{1}}, \dots, g^t_{\phi_{M}}\}$. For ease of notation, we represent the complete binary decoder by $g^b_{\phi}$ and the ternary decoder by  $g^t_{\phi}$. The main limitation of BinTurboAE and TernTurboAE is that they do not perform as well as the real-valued TurboAE. But in those applications where degradation in performance is acceptable at the cost of reduced computation and energy efficiency, BinTurboAE or TernTurboAE can be deployed at the Edge devices. As the performance of BinTurboAE is not as good as the real counterpart, each of these can be thought of as a single weak learner. Instead of relying on a single weak learner, we further propose to \textbf{\textit{ensemble}} a set of weak learners' outcomes to enable a performance that is as good as that of a real-valued network however with much lower complexity and memory requirement.
\begin{figure}[ht]
    \centering
    \caption{Architecture of the decoder of (Bin/Tern)TurboAE-Bag: the final estimate $\hat{u}$ is the aggregate of $B=4$ weak learners. $g_{\phi}^{v,i}=g_{\phi}^{b}$ for BinTurboAE and $g_{\phi}^{v,i}=g_{\phi}^{t}$ for TernTurboAE.}
    \label{fig:EnsembledAE}
\newcommand{\arrowIn}{
\tikz \draw[-stealth] (-1pt,0) -- (1pt,0);
}

\begin{tikzpicture}
 
 \draw[red,dashed] (0.25,1.6) rectangle (1.75,3.6) node[above] at (1,3.6) {Encoder} ;
 \draw[red,dashed] (4,0.25) rectangle (8.4,5.1) node[above] at (6.2,5.1) {Decoder} ;
 
 \draw (0,2.6) -- (0.5,2.6) node [pos=0.25,above] {\textbf{u}} node[pos=1]{\arrowIn};
 \draw (0.5,1.85) rectangle (1.5,3.35) node[pos=0.5] {$f_\theta$};
 
 \draw (1.5,3.1) -- (2,3.1) node [pos=0.5,above] {$\mathbf{x_1}$} node[pos=1]{\arrowIn};
 \draw (1.5,2.6) -- (2,2.6) node [pos=0.5,above] {$\mathbf{x_2}$} node[pos=1]{\arrowIn};
 \draw (1.5,2.1) -- (2,2.1) node [pos=0.5,above] {$\mathbf{x_3}$} node[pos=1]{\arrowIn};
 
 \draw (2,1.85) rectangle (3.5,3.35) node[pos=0.5] {AWGN};
 \draw (3.5,3.1) -- (4,3.1) node [pos=0.5,above] {$\mathbf{z_1}$} node[pos=1]{\arrowIn};
 \draw (3.5,2.6) -- (4,2.6) node [pos=0.5,above] {$\mathbf{z_2}$} node[pos=1]{\arrowIn};
 \draw (3.5,2.1) -- (4,2.1) node [pos=0.5,above] {$\mathbf{z_3}$} node[pos=1]{\arrowIn};
 

 \draw (4.5,4.55) -- (5,4.55) node [pos=-0.5] {$\mathbf{z_1}$} node[pos=1]{\arrowIn};
 \draw (4.5,4.3) -- (5,4.3) node [pos=-0.5] {$\mathbf{z_2}$} node[pos=1]{\arrowIn};
 \draw (4.5,4.05) -- (5,4.05) node [pos=-0.5] {$\mathbf{z_3}$} node[pos=1]{\arrowIn};
 \draw (5,3.8) rectangle (6,4.8) node[pos=0.5] {$g_\phi^{v,1}$};
 \draw (6,4.3) -- (6.5,4.3) node [pos=0.75,above] {$\hat{\mathbf{u}}^1$} node[pos=1]{\arrowIn};
 
 \draw (4.5,3.45) -- (5,3.45) node [pos=-0.5] {$\mathbf{z_1}$} node[pos=1]{\arrowIn};
 \draw (4.5,3.2) -- (5,3.2) node [pos=-0.5] {$\mathbf{z_2}$} node[pos=1]{\arrowIn};
 \draw (4.5,2.95) -- (5,2.95) node [pos=-0.5] {$\mathbf{z_3}$} node[pos=1]{\arrowIn};
 \draw (5,2.7) rectangle (6,3.7) node[pos=0.5] {$g_\phi^{v,2}$};
 \draw (6,3.2) -- (6.5,3.2) node [pos=0.75,above] {$\hat{\mathbf{u}}^2$} node[pos=1]{\arrowIn};
 
 \draw (4.5,2.35) -- (5,2.35) node [pos=-0.5] {$\mathbf{z_1}$} node[pos=1]{\arrowIn};
 \draw (4.5,2.1) -- (5,2.1) node [pos=-0.5] {$\mathbf{z_2}$} node[pos=1]{\arrowIn};
 \draw (4.5,1.85) -- (5,1.85) node [pos=-0.5] {$\mathbf{z_3}$} node[pos=1]{\arrowIn};
 \draw (5,1.6) rectangle (6,2.6) node[pos=0.5] {$g_\phi^{v,3}$};
 \draw (6,2.1) -- (6.5,2.1) node [pos=0.75,above] {$\hat{\mathbf{u}}^3$} node[pos=1]{\arrowIn};
 
 \draw (4.5,1.25) -- (5,1.25) node [pos=-0.5] {$\mathbf{z_1}$} node[pos=1]{\arrowIn};
 \draw (4.5,1) -- (5,1) node [pos=-0.5] {$\mathbf{z_2}$} node[pos=1]{\arrowIn};
 \draw (4.5,0.75) -- (5,0.75) node [pos=-0.5] {$\mathbf{z_3}$} node[pos=1]{\arrowIn};
 \draw (5,0.5) rectangle (6,1.5) node[pos=0.5] {$g_\phi^{v,4}$};
 \draw (6,1) -- (6.5,1) node [pos=0.75,above] {$\hat{\mathbf{u}}^4$} node[pos=1]{\arrowIn};

 \draw (6.5,4.3) -- (6.8,3.1) -- (7,3.1)  ;
 \draw (6.5,3.2) -- (6.75,2.85) -- (7,2.85);
 \draw (6.5,2.1) -- (6.75,2.35) -- (7,2.35);
 \draw (6.5,1) -- (6.75,2.1) -- (7,2.1) ;
 \draw (7,1.85) rectangle (8.2,3.35) node[pos=0.5] {$\frac{\sum_{b=1}^{4}\hat{\mathbf{u}}^b}{4}$} ;
 
 \draw (8.2,2.6) -- (8.7,2.6) node [pos=0.75,above] {$\hat{\mathbf{u}}$} node[pos=1]{\arrowIn};

 \end{tikzpicture}
\end{figure}
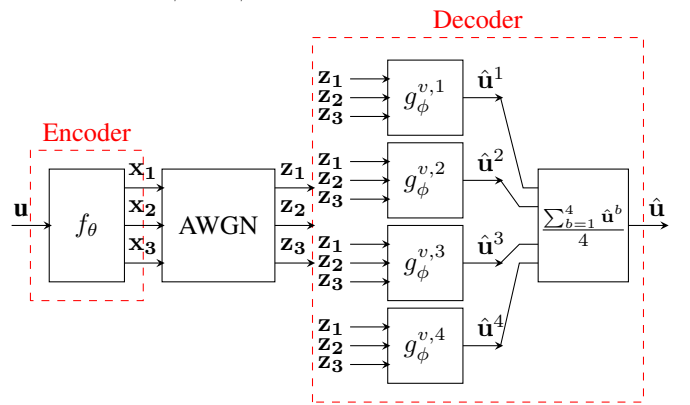

\subsection{Proposed Ensembled binary TurboAE}
Considering each decoder $g^b_{\phi}$ a weak learner, $B$ such weak learners are trained separately with the complete dataset. The idea of ``ensemble" is to get opinions from all these weak learners to arrive at a better prediction. One of the many ways the weak learners can be ensembled is \textit{Bagging} \cite{zhu2019binary}. In this work, we have proposed to ensemble $B$ BinTurboAEs with the Bagging method and denote this proposed method as \textit{BinTurboAE-Bag}. The same with TernTurboAE is called \textit{TernTurboAE-Bag}. Bagging is used in machine learning to improve stability and accuracy and to reduce variance. In Bagging method, the decisions from each one of these $B$ BinTurboAEs ($\{\hat{u}^1,\dots,\hat{u}^B\}$) are averaged to get the final prediction $\hat{\mathbf{u}}=\frac{1}{B}\sum_{b=1}^B \hat{\mathbf{u}}^b$ as shown in Fig. \ref{fig:EnsembledAE}.


\section{Experiments}
To validate the usefulness of the proposed compression techniques, we consider the setting used in \cite{NEURIPS2019_228499b5} to train the encoder and decoder networks. A large batch size, preferably greater than or equal to $500$, is used to average the channel noise effects. We train the encoder and decoder separately to avoid any possible local optima. BinTurboAE and TernTurboAE need a smaller learning rate than the real-valued TurboAE. Hence we reduced the learning rate by 10 times whenever the validation loss gets saturated for higher training epochs. The hyper-parameters used in our experiment are shown in Table \ref{tab:table1}. 

\begin{table}[h!]
  \begin{center}
    \caption{Hyper-parameters of TurboAE}
    \label{tab:table1}
    \begin{tabular}{ |p{2.5cm}| p{5cm}|  }
 \hline
  Loss & Binary Cross-Entropy (BCE)\\
  Encoder & 2 layers 1D-CNN, kernel size 5, 100 filters for each learnable encoding block \\
  Decoder & 5 layers 1D-CNN, kernel size 5, 100 filters for each learnable decoding block \\
  Decoder Iterations & 6 \\
  Info Feature Size F & 5 \\
  Batch Size & 500 \\
  Optimizer & Adam \\
  Learning Rate & initially 0.0001 and reduced by 10 when test loss saturates for more number of epochs\\
  Block Length K & 100 \\
  Number of Epochs & 800 \\
 \hline
\end{tabular}
\end{center}
\end{table}
\begin{table*}[h!]
  \begin{center}
    \caption{Savings vs performances at the edge device}
    \label{tab:table2}
    \begin{tabular}{ |p{3.8cm}| p{2.5cm}| p{2.5cm}| p{1.5cm}| p{2.5cm}| }
 \hline
 \textbf{Model} & \textbf{Memory savings} & \textbf{Computation} &\textbf{Speed up} & \textbf{BER at SNR $0$ dB} \\
 \hline
 Full precision DNN & 1x & $\simeq 4e8$ FLOPs & 1x & $1e-2$\\
 QuantTurboAE ($q=4$) & $\simeq(64/q)$x$=16$x & $\simeq 4e8$ FLOPs & 1x & $6e-2$ (q=4)\\
 BinTurboAE & $\simeq 64$x & $\simeq 4e8$ xnor-count & $64$x & $1e-1$\\
 TernTurboAE & $\simeq 64$x & $\simeq 4e8$ xnor-count & $64$x & $6e-2$\\
 (Bin/Tern)TurboAE-bag ($B=4$) & $\simeq (64/B)$x$=16$x &$\simeq$ $16e8$ xnor-count & $64$x & $2e-3$\\  
 \hline
\end{tabular}
\end{center}
\end{table*}
\begin{figure}[!t]
    \centering
    \caption{Performance of Binary and Ternary networks compared to the quantized and real valued TurboAE}
    \includegraphics[scale=0.63]{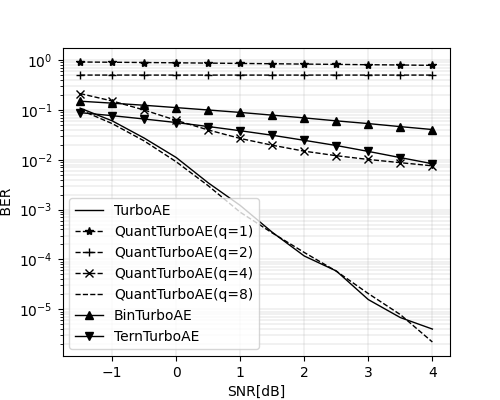}
    \label{fig:bnntnn}
\end{figure}

\begin{figure}[!t]
    \centering
    \caption{Performances of Ensembled, binary and ternary TurboAE}
    \includegraphics[scale=0.63]{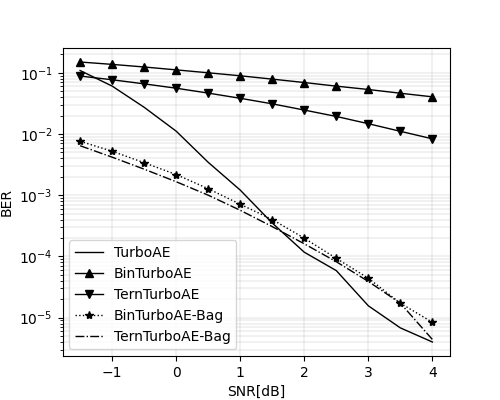}
    \label{fig:ensemble}
\end{figure}
\subsection{Results}
We provide results showing performance in terms of BER vs SNR of the proposed BinTurboAE and TernTurboAE and compare them with QuantTurboAE, the quantized TurboAE to $q$ levels \textbf{after the training}. For QuantTurboAE, the parameters of the trained TurboAE are quantized to different levels i.e. $8$-bit, $4$-bit, $2$-bit, and $1$-bit. The saving in memory is $8$, $16$, $32$, and $64$ times respectively compared to the real-valued TurboAE network as shown in Table. \ref{tab:table2}. QuantTurboAE does not offer any saving in computation unlike our proposed method. The $8$-bit quantization after the training performs as well as the original TurboAE. But the $2$-bit and $1$-bit quantizations have very poor performance as shown in Fig. \ref{fig:bnntnn}. But instead of quantization after the training, if the network is trained with $1$-bit quantization like the BinTurboAE, the network outperforms 2-bit and 1-bit QuantTurboAEs. The Ternary network improves the BER performance even more by $0.5$dB and performs similar to QuantTurboAE ($q=4$) which uses $4$ bits to store each parameter whereas the TernTurboAE uses only $1$ bit. Therefore, compared to the real-valued TurboAE, both the binary and the ternary variants save the memory requirement by about $64$ times and the computations by converting all the floating-point computations to xnor and pop-count operations at the decoder side. The performance gap between the proposed methods and TurboAE still exists and needs one's attention. To close this gap, $B=4$ such BinTurboAE as weak learners are ensembled and its performance is shown in Fig. \ref{fig:ensemble}. 

The ensemble of just $B=4$ BinTurboAEs implemented with the bagging method performs much better than that of a single BinTurboAE. The BinTurboAE-Bag even outperforms the real network in the low SNR region by almost $1$ dB. The performance of TernTurboAE-Bag is slightly better than BinTurboAE-Bag as shown in the figure. In the high SNR region, the BinTurboAE-Bag performs close to the real TurboAE. This result is significant as the BinTurboAE-Bag saves a lot in terms of the memory requirement (about $64/B$ times) and the number of computations (FLOPs are replaced with xnor-count) at the edge device end without compromising the BER performance.

\subsection{Computation and memory savings at the edge devices}
Decoding usually happens at the edge device. In real TurboAE, the iterative decoder has a huge number of parameters that take up a lot of memory. It also involves floating-point operations thus making the computations slow at the edge devices. Our main goal is then to reduce the memory requirement and computations at the decoder side of the TurboAE so that the proposed decoders are suitable for deployment at the edge. The savings for each of the proposed techniques are shown in Table. \ref{tab:table2}. BinTurboAE and TernTurboAE take up memory $64$ times lesser than the real-valued TurboAE. BinTurboAE-Bag takes a memory $B$ times of the BinTuboAE.

The number of FLOPs in the decoder of the real TurboAE at the edge devices is about $4e8$. Even though the memory savings in $q$ bit Quantized network would be around $(64/q)$ times the real network's requirement, QuantTurboAE and TurboAE do not speed up the computations as the computations are still in $64$ bit. As the Binary, Ternary and the Ensembled TurboAEs convert all the $4e8$ floating-point operations to only bitwise operations, the computations are extremely fast with much lower power consumption. When $64$ bitwise operations are performed in a single clock cycle, then the binary and ternary networks are $64$ times faster thus leading to very low latency when compared with the real TurboAE network. Even though the computation in BinTurboAE-Bag is $B$ times of the BinTurboAE, if parallel processing is available at edge, then BinTurboAE-Bag can be equally fast like BinTurboAE.

    \section{Conclusion}
In summary, we propose BinTurboAE and TernTurboAE intending to deploy the end-to-end channel coding in the targeted low-power edge devices by reducing the memory requirement and the computations by nearly $64$ times at the cost of acceptable performance degradation. We then propose BinTurboAE-bag and TernTurboAE-bag to improve the performance offered by a single BinTurboAE or single TernTurboAE respectively and achieve the performance close to the real network. The ensembled technique implemented with four such weak learners is shown to consume $16$ times less memory and computing power than the real-valued TurboAE with nearly similar performance.

 	\bibliographystyle{IEEEtran}
    \typeout{}\bibliography{library}
\end{document}